\documentclass[a4paper,10pt,twocolumn,french,twoside]{article}
\usepackage{latexsym}
\usepackage{a4wide}
\usepackage[dvips]{graphicx}
\usepackage{graphics}
\usepackage{amsmath,amsthm}
\usepackage{amsfonts,amssymb}
\usepackage[latin1]{inputenc}
\usepackage[T1]{fontenc}
\usepackage{indentfirst}
\usepackage[french,english]{babel}
\usepackage{cite}
\usepackage{geometry}
\usepackage{fancyhdr}
\usepackage{ae}
\begin{document}
\lhead{Bernard Martinie}
\chead{\thepage}
\begin{onecolumn}
\selectlanguage{french}
\title{Simulation Monte-Carlo du mod\`ele de Hubbard \`a deux dimensions}
\maketitle
\author{Bernard Martinie}
\\
\newcommand\institute[1]{{#1}}
\institute{D\'epartement de Physique, UFR Sciences et Techniques, Parc Grandmont, 37200 Tours, France}
\\
\newcommand\email[1]{{#1}}
\email {martinie@univ-tours.fr}\\
\begin{abstract}
Les r\'esultats de simulations Monte-Carlo du mod\`ele de Hubbard \`a deux dimensions \`a demi-remplissage sont pr\'esent\'es. La m\'ethode de simulation utilis\'ee est celle propos\'ee par Suzuki et al, et Hirsch et al.. Les \'etats g\'en\'er\'es par cette m\'ethode sont des \'etats de base dans la repr\'esentation en nombres d'occupation bas\'ee sur les \'etats de Wannier localis\'es autour de chaque site du r\'eseau carr\'e. Les configurations des fermions sur le r\'eseau 2D r\'eel peuvent \^etre observ\'ees. Un facteur d'antiferromagn\'etisme a pu \^etre d\'efini et calcul\'e pour chaque temp\'erature. Les courbes d'\'energie, de chaleur sp\'ecifique, de conductivit\'e et de facteur d'antiferromagn\'etisme en fonction de la temp\'erature sont pr\'esent\'ees pour diff\'erentes valeurs de l'interaction r\'epulsive coulombienne sur site $U$ (avec $t=1$). Pour les faibles valeurs de $U$ le mod\`ele pr\'esente une transition m\'etal-isolant \`a basse temp\'erature. Cette transition correspond \`a une transition paramagn\'etique-ferromagn\'etique du premier ordre. En effet, pour ces valeurs de l'interaction, les courbes d'\'energie comportent un d\'ecalage qui est caract\'eristique d'une transition du premier ordre. De m\^eme, les courbes de conductivit\'e font appara\^itre un ph\'enom\`ene d'hyst\'er\'esis qui confirme la nature de la transition. Il appara\^it un changement de comportement pour $U/t\sim3.5$. La transition change de nature. Pour les valeurs $U>3.5$ le passage ferromagn\'etique-paramagn\'etique subsiste toujours mais sans influence sur l'\'energie et la chaleur sp\'ecifique. La transistion m\'etal-isolant n'existe plus, la conductivit\'e restant faible. Les r\'eseaux d'isothermes des diff\'erentes grandeurs , en fonction de $U/t$, font appara\^itre une transition qui semble correspondre \`a la transition m\'etal-isolant de Mott. Ces r\'esultats permettent de tracer un diagramme de phase avec deux lignes de transition du premier ordre.
\end{abstract}
\selectlanguage{english}
\begin{abstract}
The Quantum Monte-Carlo simulations of the two-dimensional Hubbard model are presented for the half filling. The method based on the direct-space proposed by Suzuki and al., and Hirsch and al. was used. The states generated by this method are basis states in occupation number representation built with Wannier states localised on each site of the square array. The configurations of fermions can be observed on the real 2D array. An antiferromagnetic factor is defined and calculated for each temperature. The curves of energy, specific heat, conducivity and antiferromagnetic factor are presented for different values of the repulsive coulombian on site interaction $U$. There is a metal-insulator transition at low temperature for small values of $U$. This transition corresponds with a paramagnetic-ferromagnetic first order transition. Indeed, for these interaction values, the energy curves show a gap which is a characteristic of a first order transition. An hysteresis phenomenon appears on the conductivity curves. There is a behaviour change for $U/t\sim3.5$. For the values $U>3.5$ there is ferromagnetic-paramagnetic change without observable effect on the energy and the specific heat. The metal-insulator transition does not exist any more, the conductivity stays very small. Isotherms of the physical quantities versus $U/t$ show a transition which seems to be the metal-insulator Mott transition. These results allow to draw a phase diagram with two first order transition lines.
\end{abstract}
\end{onecolumn}
\nopagebreak
\begin{twocolumn}
\selectlanguage{french}
\section{Introduction}
\label{intro}
Le mod\`ele de Hubbard \`a deux dimensions est certainement le mod\`ele le plus simple pour d\'ecrire le comportement d'un syst\`eme de fermions fortement corr\'el\'es. Il semble suffisant pour expliquer la transition de phase m\'etal-isolant pr\'evue par Mott. La solution exacte de ce mod\`ele n'\'etant pas connue il a \'et\'e \'etudi\'e avec de nombreuses  m\'ethodes analytiques et num\'eriques \cite{Castellani,Hirsch3,pruschke,vries,Moeller,Georges,Georges-2,Duffy,Mancini,Rozenberg,Paiva,Moukouri,Dolcini,Aryanpour,Vilk,Allen,Kyung}.\\
Un grand nombre de ces r\'esultats approch\'es ont \'et\'e obtenus par des simulations num\'eriques r\'ealis\'ees avec la m\'ethode de Monte-Carlo quantique appel\'ee m\'ethode du d\'eterminant. Cette m\'ethode bas\'ee sur la formule de Trotter-Suzuki utilise la transformation de Hubbard-Stratonovich \cite{Blanckenbecler1,White1}.\\
Dans cet article nous pr\'esentons les r\'esultats de simulation obtenus par une autre  m\'ethode de Monte-Carlo, propos\'ee par Suzuki \cite{Suzuki1,Suzuki2} et Hirsch \cite{Hirsch1,Hirsch2}. Cette m\'ethode permet de g\'en\'erer des configurations des fermions dans l'espace r\'eel ce qui n'est pas le cas pour les autres m\'ethodes num\'eriques. Cette nouvelle m\'ethode, qui est la version 2D de la m\'ethode des ``\textit{world lines}, nous a permis de retrouver le comportement g\'en\'eral pr\'evu par les autres m\'ethodes et d'obtenir des r\'esultats compl\'ementaires.\\
Cet article est organis\'e comme il suit:
\begin{itemize}
\item dans la section 2 nous pr\'esentons les param\`etres du mod\`ele simul\'e, 
\item la m\'ethode est rappel\'ee dans la section 3,
\item les r\'esultats sont pr\'esent\'es dans la section 4,
\item nous analysons les r\'esultats dans la section 5,
\item la conclusion est donn\'ee dans la derni\`ere section.
\end{itemize}
\section{Mod\`ele de Hubbard}
\label{sec:1}
 Le mod\`ele \'etudi\'e est un r\'eseau carr\'e. Les \'etats monoparticulaires utilis\'es pour construire les \'etats de base $\vert\Psi_{i}\rangle$ dans la repr\'esentation en nombre d'occupation sont les \'etats de Wannier localis\'es sur les sites. L'hamiltonien de Hubbard est 
\begin{equation}
\label{hamiltonian1} 
H =-t\sum_{\left\langle i,j\right\rangle,\sigma }\left( c^{\dagger}_{j,\sigma}c_{i,\sigma}+hc\right) +U\sum_{i}n_{i\downarrow}n_{i\uparrow}
\end{equation}
o\`u les op\'erateurs $c^{\dagger}_{i,\sigma}$ et $c_{i,\sigma}$ sont les op\'erateurs de cr\'eation et d'annihilation d'un fermion de spin $\sigma$ sur le site $i$. $\langle i,j\rangle$ indique que la sommation porte sur les premiers voisins.
\section{M\'ethode num\'erique}
La m\'ethode de simulation est celle d\'evelopp\'ee dans la r\'ef\'erence \cite{martinie}. Son principe est pr\'esent\'e dans les r\'ef\'erences \cite{Suzuki2,Hirsch1,Hirsch2}. Nous en rappelons les principales caract\'eristiques.
\subsection{Principe de la m\'ethode de simulation}
 Dans l'ensemble canonique la valeur moyenne d'une observable, O, est donn\'ee par
\begin{equation}
\label{valeurmoyen}
\left\langle O\right\rangle =tr \left( DO\right) 
\end{equation} 
o\`u $D$ est l'op\'erateur densit\'e et $Z$ est la fonction de partition.
\begin{eqnarray}
\label{opdensity1}
D=\frac{e^{-\beta H}}{Z}\\
\label{partfunct1} 
Z =tr\left( e^{-\beta H}\right) 
\end{eqnarray} 
$\beta=1/k_{B}T$ est l'inverse de la  temperature. L'interaction de saut ''t`` introduit des \'el\'ements non-diagonaux, en cons\'equence les \'etats propres du hamiltonien $H$  (Eq.(\ref{hamiltonian1})), ne sont pas les \'etats de base. Aussi le calcul de la trace dans les \'equations Eqs.(\ref{valeurmoyen}) et (\ref{partfunct1}) est tr\`es probl\'ematique. La m\'ethode propos\'ee par Suzuki et Hirsch permet de contourner cette difficult\'e \cite{Suzuki1,Hirsch1}.\\
L'hamiltonien du syst\`eme est d\'ecompos\'e en plusieurs sous-hamiltoniens $H_{r}$. A cause des relations d'anticommutation certains de ces sous-hamiltoniens ne commutent pas. Cette d\'ecomposition n'est pas totalement arbitraire, les sous-hamiltoniens sont choisis tels qu'ils peuvent \^etre eux-m\^emes d\'ecompos\'es en plusieurs sous-syst\`emes sans site commun dont les hamiltoniens $K_{r,k}$ commutent.
\begin{eqnarray}
H&=&\sum_{r=1}^{p}H_{r} \hspace{1cm} \left[ H_{r},H_{r^{\prime}}\right]\neq0\hspace{1cm}  r\neq r^{\prime}\\
\label{subhamilton1} 
H_{r}&=&\sum_{k=1}^{m_{r}}K_{r,k} \hspace{0.5cm} \left[ K_{r,k},K_{r,k^{\prime}} \right] =0 \hspace{1cm}  \forall k,k^{\prime}
\end{eqnarray} 
Les d\'ecompositions pr\'ec\'edentes sont r\'ealis\'ees dans l'intention d'utiliser la formule de Trotter, Eq.(\ref{Trotter}), qui permet de contourner le probl\`eme de la non-commutativit\'e des sous-hamiltoniens $H_{r}$.
\begin{equation}
\label{Trotter} 
\exp\left( -\beta\sum_{r=1}^{p}H_{r}\right) =\lim_{n\rightarrow\infty}\left[ \prod_{r=1}^{p}\left[  \exp\left( -\frac{\beta}{n}H_{r}\right) \right] \right] ^{n}
\end{equation} 
En utilisant la formule de Trotter, la fonction de partition $Z$ s'\'ecrit
\begin{equation}
\label{partfunct2} 
Z=\lim_{n\rightarrow\infty}Z_{n}
\end{equation} 
o\`u $Z_{n}$ est un approximant de la fonction de partition 
\begin{equation}
\label{partfunct3} 
Z_{n}=tr\left\lbrace \left[ \prod_{r=1}^{p}\left[  \exp \left( -\frac{\beta}{n}H_{r}\right) \right] \right] ^{n}\right\rbrace 
\end{equation} 
En ins\'erant $np$ ensembles complets d'\'etats de base entre les op\'erateurs, l'approximant  $Z_{n}$ devient
\begin{eqnarray}
\label{partfunct4} 
 Z_{n}=\sum_{\left\lbrace \left[  \Psi_{\alpha} \right]  \right\rbrace } \langle \Psi_{0} \vert \exp \left( -\frac{\beta}{n} H_{p}\right)  \vert \Psi_{np-1} \rangle \nonumber \\ \langle \Psi_{np-1} \vert \exp \left( -\frac{\beta}{n} H_{p-1} \right) \vert \Psi_{np-2} \rangle  \nonumber \\ \ldots \langle \Psi_{1}\vert \exp \left( -\frac{\beta}{n}H_{1}\right) \vert\Psi_{0}\rangle
\end{eqnarray} 
$\left[  \Psi_{\alpha}\right] $ repr\'esente la configuration des $np$ \'etats $\vert\Psi_{j}\rangle$, et peut \^etre consid\'er\'e comme l'\'etat d'un syst\`eme classique de dimension $\left( d+1\right) $, o\`u $d$ est la dimension du syst\`eme quantique \'etudi\'e.  $\left\lbrace \left[  \Psi_{\alpha} \right]  \right\rbrace $ indique que la somme porte sur toutes les configurations possibles. Ceci est \'equivalent \`a diviser le temps imaginaire $\tau$, tel que $0\leq\tau\leq\beta$, en $n$ intervalles de dur\'ee $\bigtriangleup\tau=\beta/n$.
On d\'etermine un approximant de l'\'energie du syst\`eme \ \`a partir de la relation
\begin{equation}
\label{energy1} 
U_{n}=-\frac{\partial}{\partial\beta}\ln Z_{n}=-\frac{1}{Z_{n}}\frac{\partial Z_{n}}{\partial \beta}
\end{equation} 
En rempla\c{c}ant $Z_{n}$ par Eq.(\ref{partfunct4}), on obtient: 
\begin{eqnarray}
\label{energy2} 
U_{n} &=& \sum_{\left\lbrace \left[  \Psi_{\alpha} \right]  \right\rbrace } P _{n}\left( \left[  \Psi_{\alpha}\right]  \right) E_{n}\left( \left[ \Psi_{\alpha}\right] \right)  \\
E_{n}\left( \left[ \Psi_{\alpha}\right] \right) &=& \sum_{j=0}^{np-1}\frac{\langle\Psi_{j+1}\vert\frac{H_{r}}{n} \exp \left( -\frac{\beta}{n}H_{r}  \right) \vert\Psi_{j}\rangle}{\langle\Psi_{j+1}\vert \exp \left( -\frac{\beta}{n}H_{r} \right) \vert\Psi_{j} \rangle}\\
\label{Pn} 
 P_{n}\left( \left[ \Psi_{\alpha}\right] \right)  &=& \frac{1}{Z_{n}} \langle \Psi_{0} \vert \exp \left( -\frac{\beta}{n} H_{p}\right)  \vert \Psi_{np-1} \rangle  \nonumber \\ 
& & {} \ldots \langle \Psi_{1}\vert \exp \left( -\frac{\beta}{n}H_{1}\right) \vert\Psi_{0}\rangle 
\end{eqnarray} 
L'indice $r$ du sous-hamiltonien qui appara\^it dans chaque \'el\'ement de matrice est fonction de l'indice de l'\'etat  $j$, tel que $r=1+\left( j \bmod p\right) $. Le calcul de la trace impose des conditions p\'eriodiques sur les \'etats, ainsi le ket $\vert\Psi_{0}\rangle$ correspond au ket $\vert\Psi_{np}\rangle$. Les facteurs $P_{n}$ v\'erifient 
\begin{equation}
 \sum_{\left\lbrace \left[ \Psi_{\alpha}\right] \right\rbrace }P _{n}\left( \left[  \Psi_{\alpha}\right]  \right)=1
\end{equation} 
On peut ainsi consid\'erer que chaque configuration $ \left[ \Psi_{\alpha}\right]  $ du syst\`eme de dimension $\left( d+1\right) $ a une \'energie $E_{n}\left( \left[ \Psi_{\alpha}\right] \right)$ et un facteur de probabilit\'e $P _{n} \left( \left[  \Psi_{\alpha}\right]  \right)$. Le calcul de la valeur moyenne de l'\'energie du syst\`eme est r\'ealis\'e en utilisant une m\'ethode de Mont\'e-Carlo telle que l'algorithme de Metropolis.\\
En utilisant la relation (\ref{subhamilton1}), la valeur de $U_{n}$ s'\'ecrit:
\begin{eqnarray}
\label{Unapproxi}
U_{n} &\approx & \frac{1}{N_{p}}\sum_{\alpha=1}^{N_{p}}E_{n,\alpha} = \left\langle E_{n,\alpha}\right\rangle \\
\label{Enapproxi}  E_{n,\alpha} &=& \frac{1}{n}\sum_{i=1}^{n}\sum_{r=1}^{p}\sum_{k=1}^{m_{r}}\frac{\langle\Psi_{j+1}\vert  K_{r,k}\exp \left( -\frac{\beta}{n}\sum_{l} K_{r,l}\right) \vert\Psi_{j}\rangle}{\langle\Psi_{j+1}\vert \exp \left( -\frac{\beta}{n}\sum_{l} K_{r,l}\right) \vert\Psi_{j}\rangle} \nonumber \\ 
\end{eqnarray} 
o\`u $N_{p}$ est le nombre de configurations retenues, $\alpha$ est l'indice des configurations qui remplace $\left[ \Psi_{\alpha}\right] $ et $i$ est l'indice des intervalles de temps. Les indices $i$, $j$ et $r$ sont reli\'es par le principe de num\'erotation des $np$ \'etats $\vert \Psi_{j}\rangle$ et v\'erifient
\begin{equation}
j=\left( i-1\right)p+r-1 
\end{equation}
La d\'ecomposition des sous-hamiltoniens $H_{r}$ en sous-syst\`emes permet de r\'ealiser une simplification importante. Cette simplification n\'ecessite une approximation.
Tous les hamiltoniens $K_{r,l}$ des sous-syst\`emes  d'un m\^eme sous-hamiltonien $H_{r}$ commutent et chaque hamiltonien $K_{r,l}$ agit seulement sur l'\'etat d'un seul sous-syst\`eme aussi chaque d\'enominateur de l'expression (\ref{Enapproxi}) peut s'\'ecrire
\begin{eqnarray}
\label{approximation1} 
\langle\Psi_{j+1}\vert \exp \left( -\frac{\beta}{n}\sum_{l=1}^{m_{r}} K_{r,l}\right) \vert\Psi_{j}\rangle \longrightarrow \nonumber \\ \prod_{l=1}^{m_{r}}\langle \varphi_{j+1,l} \vert  \exp \left( -\frac{\beta}{n}K_{r,l}\right)  \vert \varphi_{j,l}\rangle
\end{eqnarray} 
o\`u l'\'etat $\vert \varphi_{j,l}\rangle$  est l'\'etat du sous-syst\`eme d'indice $l$ du sous-hamiltonien d'indice $r$ de l'intervalle de temps d'indice $i$. Le symbole  $\longrightarrow$ signifie que l'expression du membre de gauche est remplac\'ee par celle du membre de droite. Cette op\'eration implique que l'espace des \'etats du syst\`eme est consid\'er\'e comme le produit tensoriel des espaces des \'etats de tous les sous-syt\`emes. Ceci n'est pas correcte pour un syst\`eme de fermions, en effet, \`a cause de l'antisym\'etrie des \'etats, les op\'erateurs cr\'eation et annihilation sont d\'efinis dans l'espace des \'etats de tout le syst\`eme. Il n'y a pas une \'egalit\'e stricte entre ces deux expressions. Implicitement ceci signifie que les nombres d'occupation des autres \'etats monoparticulaires ne sont pas pris en compte.\\
Etant donn\'e la forme particuli\`ere des hamiltoniens, cette approximation peut, selon les valeurs des nombres d'occupation des \'etats monoparticulaires, changer le signe de certains \'el\'ements non-diagonaux des $K_{r,l}$.\\
 Si le nombre de types de sous-syst\`emes est faible, cette simplification permet de r\'eduire consid\'erablement les calculs en diminuant les dimensions de l'espace des \'etats o\`u sont men\'e ces calculs. En effet, la diagonalisation du hamiltonien $H$ de tout le syst\`eme dans l'espace des \'etats de dimension $2^{N}$, o\`u $N$ est le nombre de sites de tout le syst\`eme, est remplac\'ee par la diagonalisation des hamiltoniens $K_{r,k}$ de quelques sous-syst\`emes, dont les espaces des \'etats sont de tr\`es petites dimensions compar\'ees \`a $2^{N}$.\\
On applique la m\^eme factorisation pour le num\'erateur de l'\'equation Eq. (\ref{Enapproxi}). Les expressions calcul\'ees sont finalement:
\begin{eqnarray}
U_{n}^{\prime}&=&\left\langle E_{n,\alpha}^{\prime} \right\rangle \\
\label{Enalphaprime}
 E_{n,\alpha}^{\prime}&=& \frac{1}{n}\sum_{i=1}^{n}\sum_{r=1}^{p}\sum_{k=1}^{m_{r}}\frac{\langle \varphi_{j+1,k}\vert  K_{r,k}\exp \left( -\frac{\beta}{n}K_{r,k}\right) \vert \varphi_{j,k}\rangle}{\langle \varphi_{j+1,k}\vert \exp \left( -\frac{\beta}{n} K_{r,k}\right) \vert \varphi_{j,k}\rangle} \nonumber \\
\end{eqnarray} 
avec
\begin{eqnarray}
\label{Pnalphaprime} 
P_{n,\alpha}^{\prime}=\frac{1}{Z^{\prime}_{n}} \prod_{i=1}^{n}   \prod_{r=1}^{p}  \prod_{k=1}^{m_{r}}\langle \varphi_{j+1,k} \vert  \exp \left( -\frac{\beta}{n}K_{r,k}\right)  \vert \varphi_{j,k}\rangle \nonumber \\ \\ 
\label{Znprime} 
Z^{\prime}_{n}=\sum \prod_{i=1}^{n}   \prod_{r=1}^{p}  \prod_{k=1}^{m_{r}}\langle \varphi_{j+1,k} \vert  \exp \left( -\frac{\beta}{n}K_{r,k}\right)  \vert \varphi_{j,k}\rangle \nonumber \\
\end{eqnarray} 
 Certains facteurs de probabilit\'e $P _{n} \left( \left[  \Psi_{\alpha}\right] \right) $ o\`u  $P_{n,\alpha}^{\prime}$ sont n\'egatifs: c'est le ``\textit{probl\`eme du signe}''. Nous avons montr\'e dans la r\'ef\'erence \cite{martinie} qu'on peut ne pas tenir compte de ce signe.\\
La chaleur sp\'ecifique est d\'etermin\'ee \`a partir de la formule 
\begin{equation}
 c=-k_{B} \beta^{2} \frac{\partial U}{\partial\beta}
\end{equation} 
o\`u $k_{B}$ est la constante de Boltzman.\\
En utilisant une m\'ethode similaire \`a celle utilis\'ee pour $U_{n}^{\prime}$ on obtient une formule qui permet de calculer la chaleur sp\'ecifique \`a partir des fluctuations
\begin{equation}
\label{specifheat1} 
\frac{C}{R} \approx \beta^{2}\left\langle \left( E_{n,\alpha}^{\prime}-U_{n}^{\prime}\right) ^{2}\right\rangle 
\end{equation} 
o\`u $C$ est la chaleur sp\'ecifique molaire et $R$ la constante des gaz parfaits.
\subsection{Conductivit\'e statique}
 La m\'ethode de simulation ne permet pas de faire une d\'etermination rigoureuse de la conductivit\'e statique. En effet, l'op\'erateur densit\'e de courant d\'efini par Scalapino and al. \cite{Scalapino1} n'est pas diagonal et ne peut pas \^etre d\'ecompos\'e en op\'erateur relatif \`a chaque sous-syst\`eme \cite{Hirsch2}. Cependant, on obtient une valeur approch\'ee de la densit\'e de courant au temps imaginaire $\tau=\frac{\beta}{np}j$, en calculant:
\begin{eqnarray}
 j_{x}\left( l;j\right)=&&\sum_{\sigma}n_{\sigma,l;j}\left( 1-n_{\sigma,l^{\prime};j}\right)  n_{\sigma,l^{\prime};j^{\prime}} \nonumber \\ &&\left( 1-n_{\sigma,l;j^{\prime}}\right)-n_{\sigma,l^{\prime};j}\left( 1-n_{\sigma,l;j}\right)\nonumber\\&& n_{\sigma,l;j^{\prime}}\left( 1-n_{\sigma,l^{\prime};j^{\prime}}\right) 
\end{eqnarray}
$j^{\prime}=j+p$ o\`u $p$ est le nombre de sous-hamiltoniens. Les coordonn\'ees des sites $l$ et $l^{\prime}$ v\'erifient $x_{l^{\prime}}=x_{l}+1$  et $y_{l^{\prime}}=y_{l}$. $n_{\sigma,l;j}$ est le nombre d'occupation de l'\'etat monoparticulaire $\vert \sigma,l\rangle$ du r\'eseau 2D d'indice $j$.
Cette formule de calcul de $j_{x}\left( l;j\right) $ correspond  \`a l'expression
\begin{equation}
\langle\psi_{j^{\prime}}\vert \sum_{\sigma}c\dagger _{\sigma,l^{\prime}}c_{\sigma,l}-c\dagger _{\sigma,l}c_{\sigma,l\prime}\vert \psi_{j}\rangle 
\end{equation} 
o\`u l'\'etat $\vert\psi_{j}\rangle$ correspond \`a l'\'etat des sites $l$ et $l^{\prime}$. La conductivit\'e calcul\'ee est: 
\begin{equation}
 \Lambda_{xx}\left( m\right) =\frac{1}{np}\sum_{j=1}^{np} \sum_{l=1}^{N} j_{x}\left( l;j\right)j_{x}\left( l;j+m\right)
\end{equation} 
o\`u $N$ est le nombre de sites du syst\`eme. La conductivit\'e statique est obtenue apr\`es une transformation de Fourier discr\`ete.
\subsection{Facteur d'antiferromagn\'etisme}
Pour chaque temp\'erature la fonction de corr\'elation de spin $c\left( l_{x},l_{y}\right) $ est calcul\'ee
\begin{equation}
\label{correlspin} 
c\left( l_{x},l_{y}\right) =\left\langle \left( \left(  n_{i,\uparrow}-n_{i,\downarrow}\right)\left( n_{i+l,\uparrow}-n_{i+l,\downarrow}\right) \right) \right\rangle  
\end{equation}
$\left\langle \cdots \right\rangle $ signifie que, pour chaque configuration retenue, on calcule la moyenne pour tous les $N$ sites de chaque r\'eseau 2D et pour tous les r\'eseaux 2D de cette configuration. La fonction finale correspond \`a la moyenne sur toutes les configurations  retenues \`a une m\^eme temp\'erature. $l$ représente le changement d'indice des sites provoqu\'e par la translation de vecteur $\left( l_{x},l_{y}\right) $ sur le r\'eseau. Le facteur d'antiferromagn\'etisme (param\`etre d'ordre de N\'eel) est d\'eduit de la fonction de corr\'elation de spin tel que
\begin{equation}
f_{A}=\frac{1}{N}\sum_{l_{x},l_{y}}\left( -1\right) ^{\left( l_{x}+l_{y}\right)} c\left( l_{x},l_{y}\right)  
\end{equation}
Dans le cas d'un ordre antiferromagn\'etique ce facteur est tel que $f_{A}=1$, par contre, il n'est pas rigoureusement nul pour une absence totale d'ordre. En effet, le terme correspondant \`a $l_{x}=l_{y}=0$ n'est pas \'elimin\'e dans notre d\'efinition de ce facteur. Cette contribution est de l'ordre de $\sim 1/N$.
\section{R\'esultats}
\label{resultats}
\subsection{Param\`etres des simulations}
Le mod\`ele \'etudi\'e est un r\'eseau carr\'e contenant $N=6\times6$ sites avec les conditions aux limites p\'eriodiques. Tous les sous-syst\`emes sont identiques et sont compos\'es de quatre sites. Ainsi la m\'ethode ne n\'ecessite la diagonalisation  que d'une seule matrice de dimension $16\times16$. Ces sous-syst\`emes sont regroup\'es dans deux sous-hamiltoniens ($p=2$). La figure \ref{fig:array2D} montre la d\'ecomposition du syst\`eme en sous-syst\`emes. Toutes les simulations ont \'et\'e r\'ealis\'ees avec huit intervalles de temps imaginaire ($n=8$). L'interaction de saut est fix\'ee \`a $t=1$. Les premiers voisins d'un site sont les quatre sites correspondant aux translations $\left( \triangle x=\pm1, \triangle y=0\right) $ et $\left( \triangle x=0, \triangle y=\pm 1\right) $. Les simulations ont \'et\'e programm\'ees pour des valeurs de l'interaction r\'epulsive coulombienne variant de  $U=0$ \`a $U=8$. La majorit\'e des simulations ont \'et\'e r\'ealis\'ees pour le demi-remplissage, c'est \`a dire dix-huit spins up et dix-huit spins down ($18\uparrow+18\downarrow$). Des simulations, pour $U=0$, ont \'et\'e men\'ees pour un seul spin up ($1\uparrow +0\downarrow$)  et dix-huit spins up sans spin down ($18\uparrow +0\downarrow$) pour comparer les r\'esultats afin de v\'erifier la coh\'erence de la m\'ethode. La plupart des simulations ont consist\'e en neuf cycles de descente-mont\'ee de temp\'erature, avec $100$ points de temp\'erature en progression g\'eom\'etrique pour chaque mont\'ee ou descente de temp\'erature. Pour chaque grandeur \'etudi\'ee trois moyennes ont \'et\'e calcul\'ees: la moyenne en mont\'ee, la moyenne en descente et la moyenne totale. Ceci a permis de mettre en \'evidence un faible ph\'enom\`ene d'hyst\'er\'esis entre la mont\'ee et la descente de temp\'erature dans une certaine plage de temp\'erature.
\begin{figure}
\resizebox{0.75\columnwidth}{!}{%
  \includegraphics{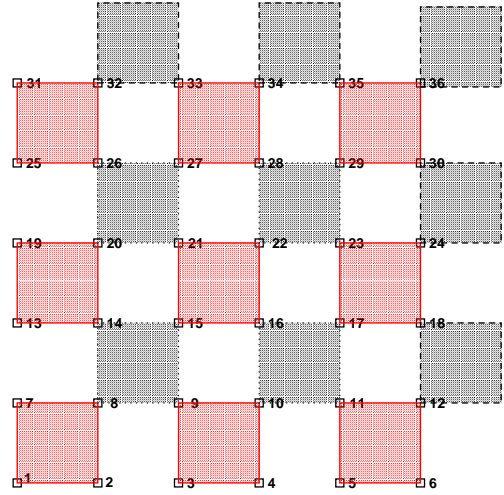}
}
\caption{(Couleur en ligne) D\'ecomposition du r\'eseau carr\'e en deux sous-hamiltoniens et en sous-syst\`emes avec les conditions aux limites p\'eriodiques. Les sous-syst\`emes gris (rouge) appartiennent au sous-hamiltonien 1, les sous-syst\`emes noirs appartiennent au sous-hamiltonien 2.}
\label{fig:array2D}
\end{figure}
\subsection{Test de la m\'ethode (U=0)}
\begin{figure}
\resizebox{0.9\columnwidth}{!}{%
  \includegraphics{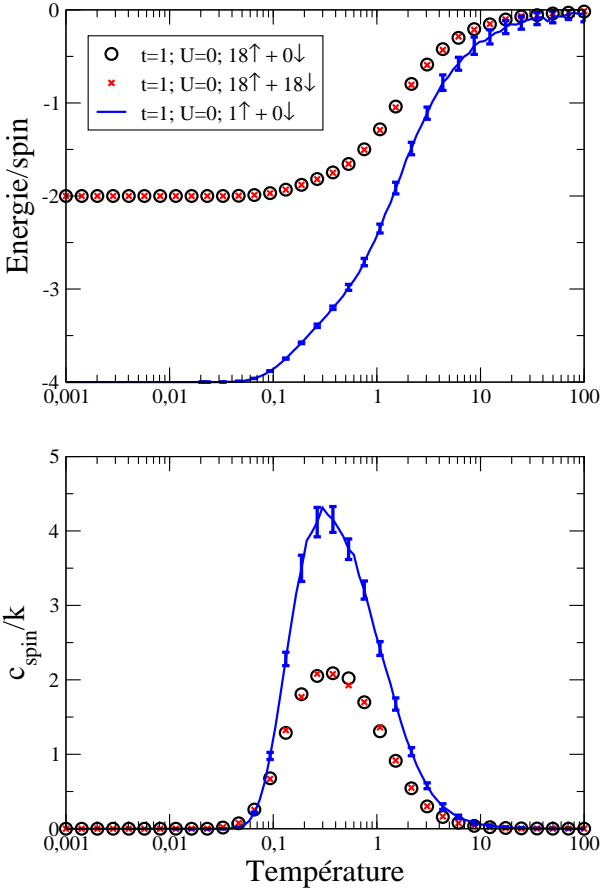}
}
\caption{(Couleur en ligne) Courbes d'\'energie et de chaleur sp\'ecifique par spin en fonction de la temp\'erature pour $U=0$.}
\label{fig:CourbeU0}
\end{figure}
\begin{figure}
\resizebox{0.9\columnwidth}{!}{%
  \includegraphics{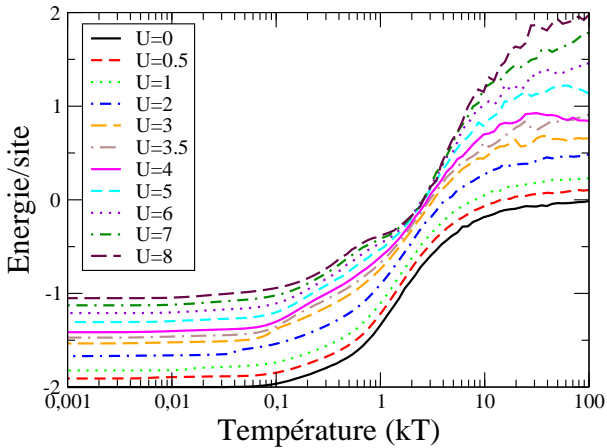}
}
\caption{(Couleur en ligne) Courbes d'\'energie par site en fonction de la temp\'erature pour diff\'erentes valeurs de $U$.}
\label{fig:Energie-1}
\end{figure}
La figure \ref{fig:CourbeU0} montre les courbes d'\'energie et de chaleur sp\'ecifique par spin en fonction de la temp\'erature pour $U=0$ et les trois remplissages suivants: $(1\uparrow +0\downarrow)$, $(18\uparrow +0\downarrow)$ et $(18\uparrow +18\downarrow)$. Pour $U=0$ le demi-remplissage $(18\uparrow +18\downarrow)$ correspond \`a deux gaz quantiques non-corr\'el\'es qui ont les m\^emes \'energies et chaleurs sp\'ecifiques. Ceci explique pourquoi les courbes sont identiques pour les remplissages  $(18\uparrow +0\downarrow)$ et $(18\uparrow +18\downarrow)$. L'\'energie et la chaleur sp\'ecifique par un seul spin sur le r\'eseau $(1\uparrow +0\downarrow)$ sont les doubles des valeurs obtenues pour les remplissages pr\'ec\'edents. Le niveau du fondamental est \'egal \`a $-4t$ comme il est pr\'evu. En effet cet \'etat correspond \`a:
\begin{eqnarray}
 \label{fondamental} 
\vert \Psi_{0}\rangle=\frac{1}{\sqrt{N}}\sum_{i=1}^{N}\vert n_{1,\uparrow}=0,\ldots ,n_{i,\uparrow}=1,n_{i,\downarrow}=0, \nonumber \\ n_{i+1,\uparrow}=0,\ldots ,n_{N,\downarrow}=0 \rangle
\end{eqnarray}
dont l'\'energie est donn\'ee par
\begin{equation}
H\vert\Psi_{0}\rangle=-t\sum_{\langle  i,j\rangle}c_{j,\uparrow}^{\dagger}c_{i,\uparrow}\vert\Psi_{0}\rangle=-4t\vert\Psi_{0}\rangle
\end{equation}
Le rapport deux entre les grandeurs calcul\'ees pour le remplissage $(1\uparrow +0\downarrow)$ et les deux autres remplissages ($(18\uparrow +0\downarrow)$ et $(18\uparrow +18\downarrow)$) peut \^etre justifi\'e en consid\'erant que pour ces deux derniers remplissages chaque spin ne peut se d\'eplacer, en moyenne, que sur deux sites voisins alors que dans le cas d'un seul spin sur le r\'eseau le spin a quatre possibilit\'es de saut.\\
Il est int\'eressant de remarquer que dans le cas d'un seul spin sur le r\'eseau le probl\`eme du signe n'existe pas, de m\^eme, l'approximation relative aux nombres d'occupation non pris en compte dans la relation (\ref{approximation1}) n'est pas utilis\'ee. Dans ce cas la m\'ethode n'utilise aucune approximation except\'e le d\'ecoupage du temps imaginaire li\'e \`a la formule de Trotter. La bonne concordance de ces r\'esultats confirme la coh\'erence de la m\'ethode de simulation.
\subsection{R\'esultats des simulations}
\begin{figure}
\resizebox{0.9\columnwidth}{!}{%
  \includegraphics{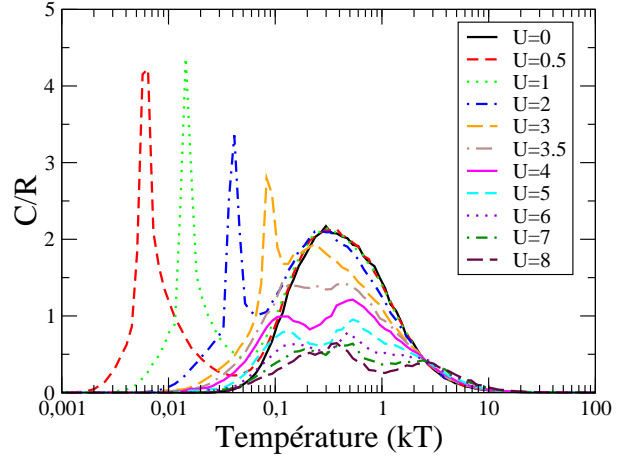}
}
\caption{(Couleur en ligne) Courbes de chaleur sp\'ecifique molaire en fonction de la temp\'erature pour diff\'erentes valeurs de $U$.}
\label{fig:Chalspec-1}
\end{figure}
\begin{figure}
\resizebox{0.9\columnwidth}{!}{%
  \includegraphics{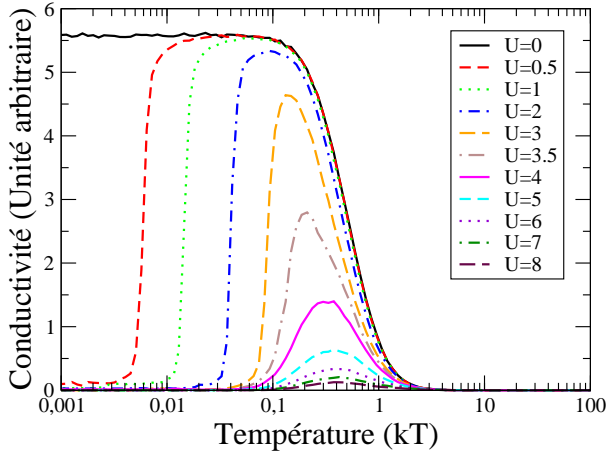}
}
\caption{(Couleur en ligne) Courbes de conductivit\'e en fonction de la temp\'erature pour diff\'erentes valeurs de $U$.}
\label{fig:Conductiv}
\end{figure}
\begin{figure}
\resizebox{0.9\columnwidth}{!}{%
  \includegraphics{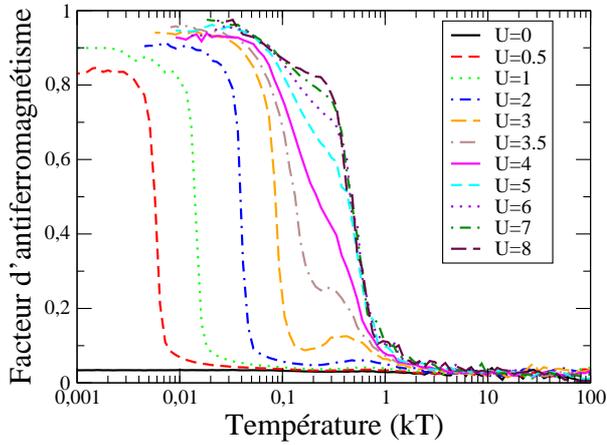}
}
\caption{(Couleur en ligne) Courbes du facteur d'antiferromagn\'etisme en fonction de la temp\'erature pour diff\'erentes valeurs de $U$.}
\label{fig:factantiferro}
\end{figure}
\begin{figure}
\resizebox{0.9\columnwidth}{!}{%
  \includegraphics{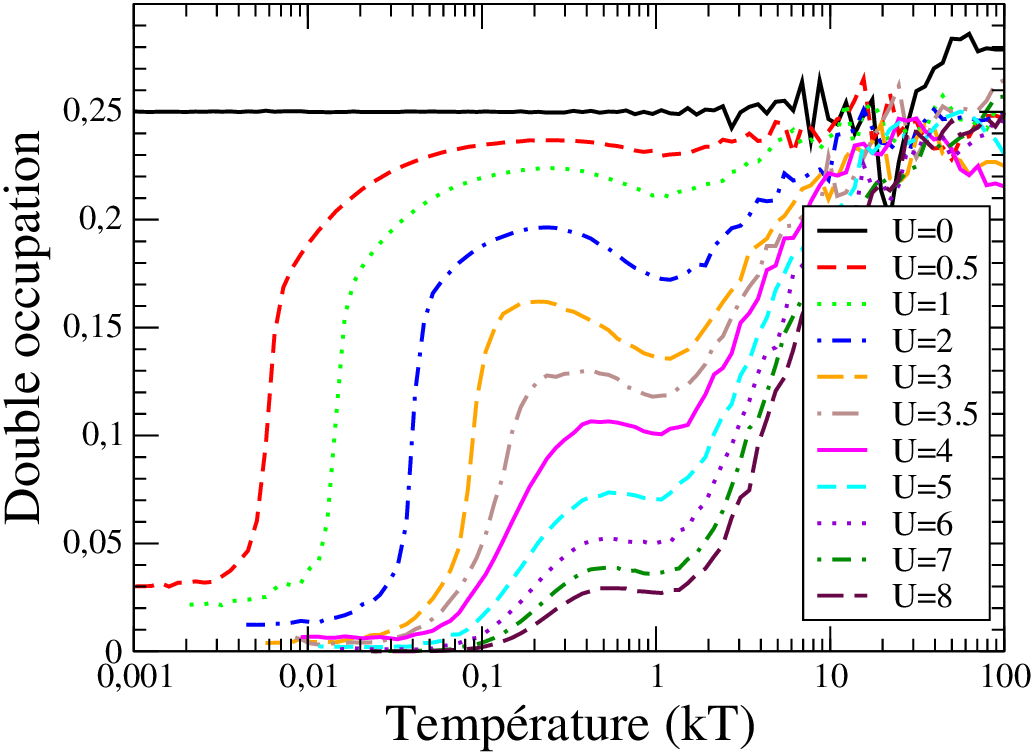}
}
\caption{(Couleur en ligne) Courbes de double ocupation en fonction de la temp\'erature pour diff\'erentes valeurs de $U$.}
\label{fig:paires}
\end{figure}
Les figures \ref{fig:Energie-1}, \ref{fig:Chalspec-1}, \ref{fig:Conductiv}, \ref{fig:factantiferro} et \ref{fig:paires} montrent les courbes moyennes d'\'energie, de chaleur sp\'ecifique, de conductivit\'e, du facteur d'antiferromagn\'etisme et de double occupation (fraction de sites occup\'es par une paire de spins $\left( \uparrow\downarrow \right) $). Ces courbes correspondent aux moyennes totales sur les mont\'ees et les descentes des neuf cycles de temp\'erature.\\
Pour $U=0$, le facteur d'antiferromagn\'etisme est $f_{A}\sim 1/N$, c'est \`a dire qu'il n'y a pas d'ordre magn\'etique quelle que soit la temp\'erature.\\
Le facteur de double occupation
\begin{equation}
\label{doubleoccup}
D=\frac{1}{N}\langle n_{i,\uparrow}n_{i,\downarrow} \rangle 
\end{equation}
peut \^etre reli\'e \`a la valeur moyenne du carr\'e de l'aimantation locale $\left\langle  m_{z} ^{2} \right\rangle $. En effet, si le moment magn\'etique de chaque particule est choisi tel que $s_{z}=\pm 1$ alors, pour chaque r\'eseau 2D:
\begin{equation}
 \sum\left( n_{i,\uparrow}-n_{i,\downarrow}\right) ^{2}=\sum \left( m_{z}\right)^{2}=\sum \vert m_{z}\vert =N_{\uparrow}+N_{\downarrow}-2\:n_{p}
\end{equation} 
o\`u $ N_{\uparrow}=\sum n_{i,\uparrow} $ et $N_{\downarrow}=\sum n_{i,\downarrow}$ sont les nombres de spins up et down, $n_{p}=\sum n_{i,\uparrow}\:n_{i,\downarrow}$ est le nombre de paires dans un r\'eseau 2D. Dans le cas pr\'esent, o\`u $N_{\uparrow}=N_{\downarrow}=N/2=18$, on obtient 
\begin{equation}
 \left\langle m_{z} ^{2}\right\rangle =1-2D
\end{equation} 
On remarque qu'\`a basse temp\'erature $ \left( kT < t \right) $, pour $U=0$, le nombre de paires est exactement $n_{p}=9$, c'est \`a dire que les spins de chaque orientation sont distribu\'es de facon totalement al\'eatoire sur le r\'eseau 2D. Ceci est en parfait accord avec le r\'esultat relatif au facteur d'antiferromagn\'etisme.\\
On remarque un point de croisement, pour $kT\simeq2.7t$, sur le r\'eseau des courbes de chaleur sp\'ecifique de la figure \ref{fig:Chalspec-1}.\\
Les figures \ref{fig:specif-lin} \`a \ref{fig:mz2-lin} repr\'esentent les courbes de chaleur sp\'ecifique, de conductivit\'e, de facteur d'antiferromagn\'etisme et de carr\'e de l'aimantation locale pour les basses temp\'eratures avec une \'echelle de temp\'erature lin\'eaire. Pour les faibles valeurs de $U$, les parties d\'ecroissantes des courbes de conductivit\'e sont assez bien approxim\'ees par des fonctions de la forme
\begin{equation}
\sigma\left( kT\right) \simeq \frac{a-b\:\left( kT\right) }{\left( kT\right) ^{2}+c}+d 
\end{equation} 
o\`u $d$ est petit.
\begin{figure}
\resizebox{0.9\columnwidth}{!}{%
  \includegraphics{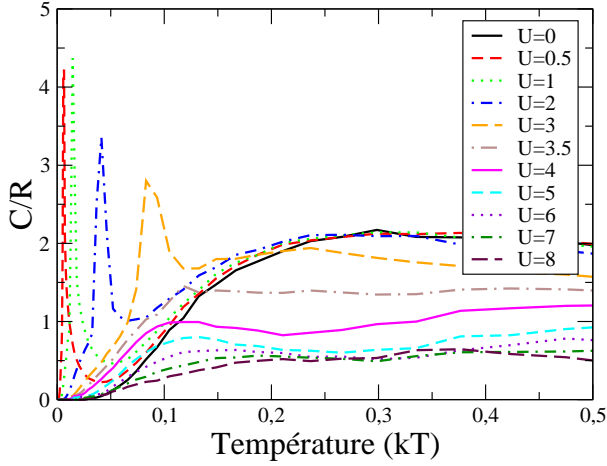}
}
\caption{(Couleur en ligne) Courbes de la chaleur sp\'ecifique pour les basses temp\'eratures.}
\label{fig:specif-lin}
\end{figure}
\begin{figure}
\resizebox{0.9\columnwidth}{!}{%
  \includegraphics{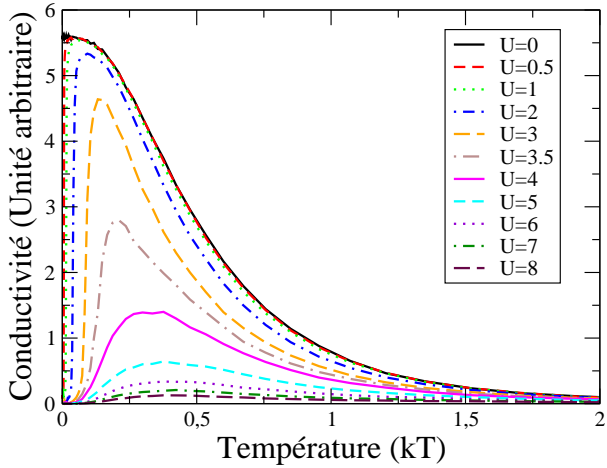}
}
\caption{(Couleur en ligne) Courbes de la conductivit\'e pour les basses temp\'eratures.}
\label{fig:conduct-lin}
\end{figure}
\begin{figure}
\resizebox{0.9\columnwidth}{!}{%
  \includegraphics{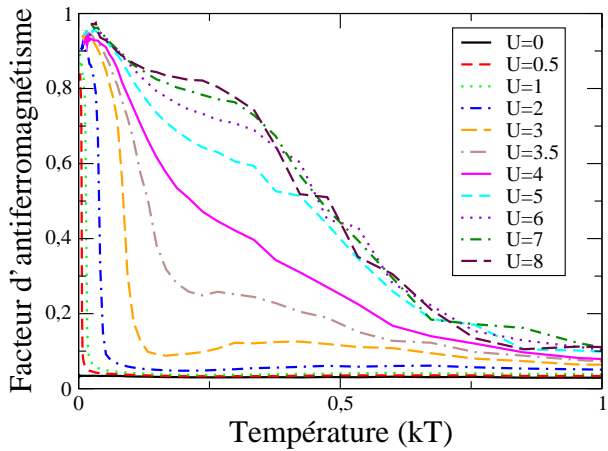}
}
\caption{(Couleur en ligne) Courbes du facteur d'antiferromagn\'etisme pour les basses temp\'eratures.}
\label{fig:factantifer-lin}
\end{figure}
\begin{figure}
\resizebox{0.9\columnwidth}{!}{%
  \includegraphics{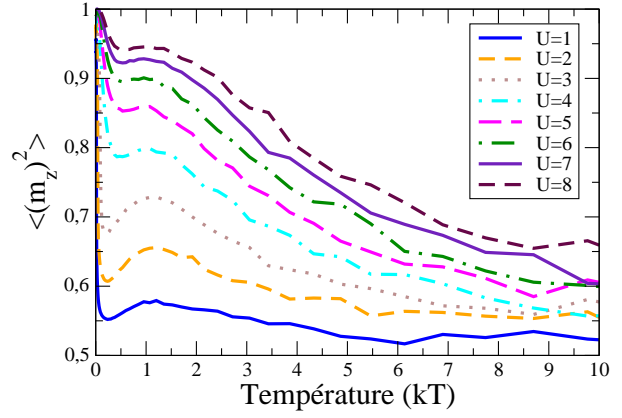}
}
\caption{(Couleur en ligne) Courbes du carr\'e de l'aimantation locale pour les basses temp\'eratures.}
\label{fig:mz2-lin}
\end{figure}
\subsection{Discontinuit\'e des courbes d'\'energie}
Pour les temp\'eratures inf\'erieures \`a environ $kT/t=3.5$ les courbes d'\'energie pr\'esentent un l\'eger d\'ecrochement qui correspond au pic de chaleur sp\'ecifique. Des simulations avec $200$ points de temp\'erature dans des plages de temp\'eratures autour ce ces d\'ecrochements ont \'et\'e r\'ealis\'ees pour $U=0.5t$, $U=1t$, $U=2t$, $U=3t$, $U=3.3t$, $U=3.5t$ et $U=4t$. Les courbes d'\'energie correspondantes (qui ne sont pas des courbes moyennes) sont pr\'esent\'ees dans les figures \ref{fig:decenergU0.5} \`a \ref{fig:decenergU3.5}. Les courbes moyennes de la figure \ref{fig:Energie-1} sont reproduites pour comparaison. Les barres d'erreur de ces courbes permettent d'\'evaluer les fluctuations. On remarque, sur toutes les courbes, que l'amplitude de ces barres d'erreur est faible en dehors de petites plages de temp\'erature. Ces plages de temp\'erature correspondent \`a des zones de fluctuation sur les courbes avec $200$ points de mesure. On constate que dans ces zones, les \'energies fluctuent entre deux valeurs limites  qui correspondent aux extrapolations des portions de courbe sans fluctuation. L'amplitude des fluctuations, qui correspond approximativement, au d\'ecrochement augmente de $U=0.5t$ \`a $U=3t$, puis diminue rapidement quand $U$ augmente. Il n'y a plus de fluctuation pour $U=3.5t$. L'amplitude maximum, estim\'ee pour $U=3t$, est d'environ $44\; 10^{-3}t$. Pour $U=3.3t$ la valeur des fluctuations d'\'energie est d'environ $14\; 10^{-3}t$, et l'extrapolation des courbes est trop ``hasardeuse'' pour pr\'esenter un int\'er\^et.
\begin{figure}
\resizebox{0.9\columnwidth}{!}{%
  \includegraphics{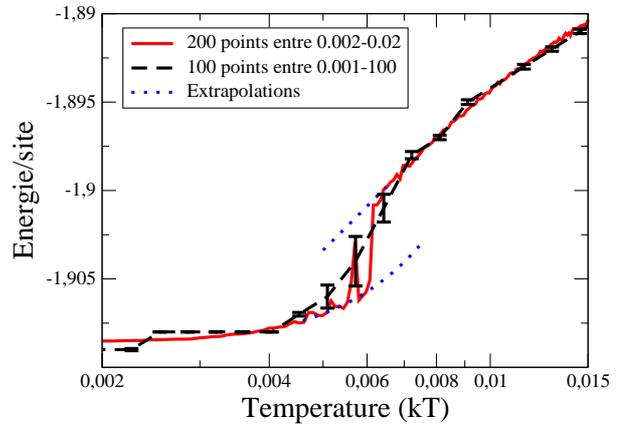}
}
\caption{(Couleur en ligne) Courbes d'\'energie pour $U=0.5t$. La courbe en tiret\'es noirs correspond \`a la courbe moyenne de la figure \ref{fig:Energie-1} avec ses barres d'erreur.}
\label{fig:decenergU0.5}
\end{figure}
\begin{figure}
\resizebox{0.9\columnwidth}{!}{%
  \includegraphics{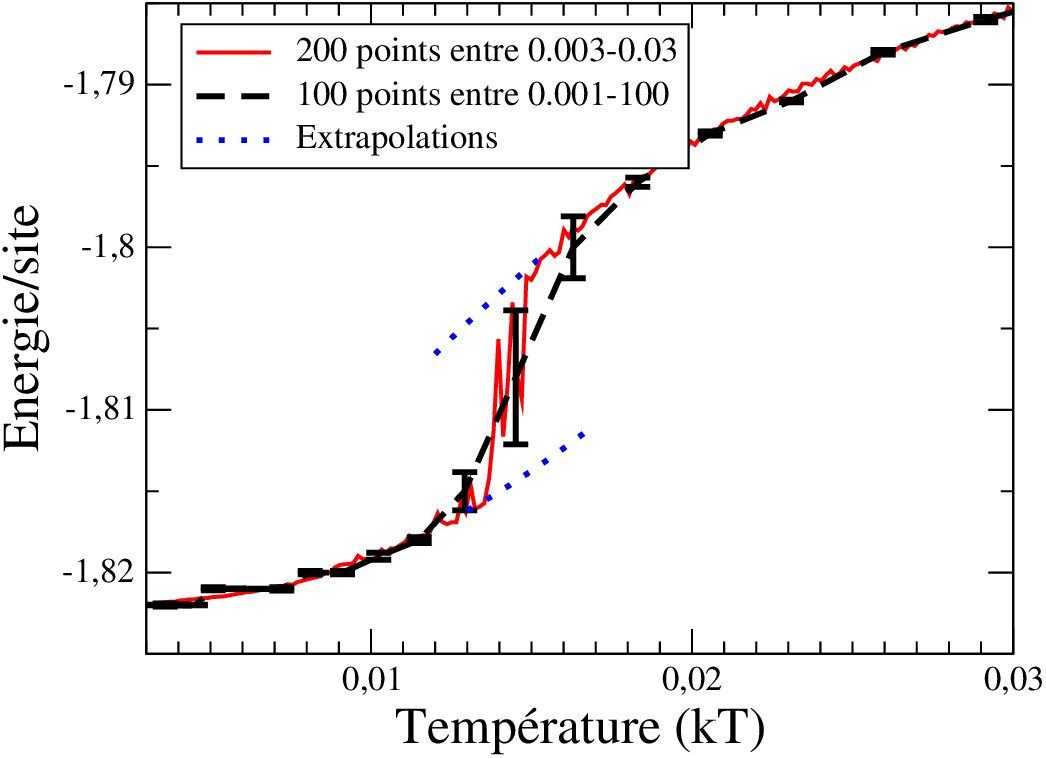}
}
\caption{(Couleur en ligne) Courbes d'\'energie pour $U=1t$. La courbe en tiret\'es noirs correspond \`a la courbe moyenne de la figure \ref{fig:Energie-1} avec ses barres d'erreur.}
\label{fig:decenergU1}
\end{figure}
\begin{figure}
\resizebox{0.9\columnwidth}{!}{%
  \includegraphics{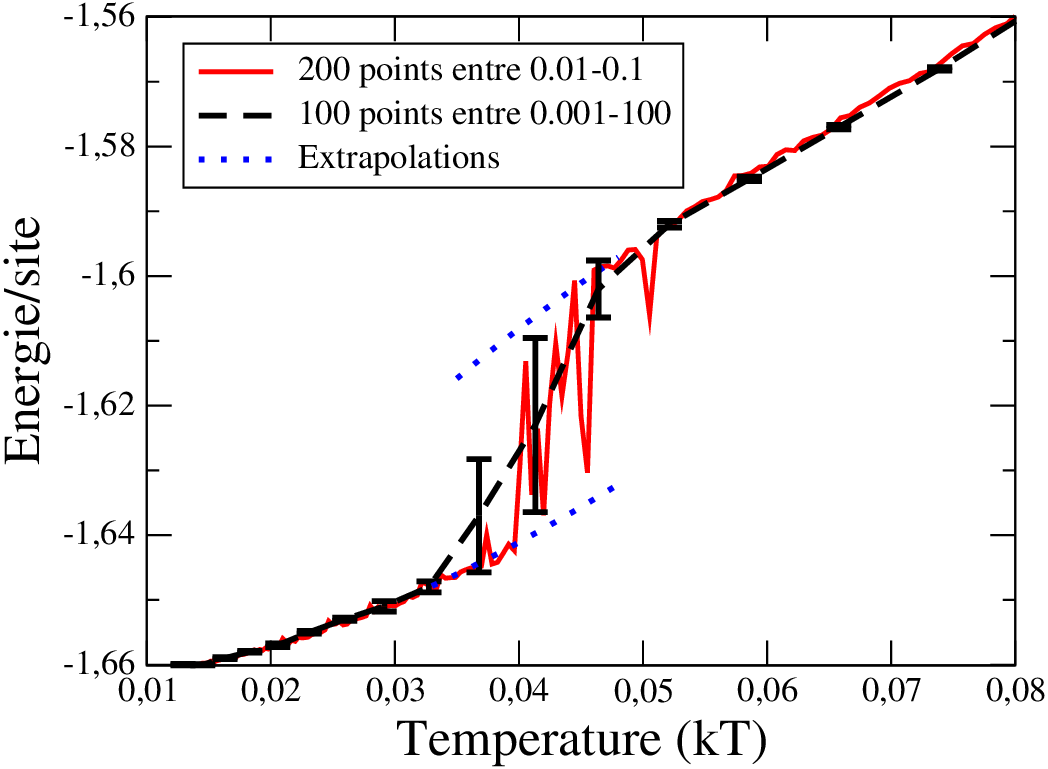}
}
\caption{(Couleur en ligne) Courbes d'\'energie pour $U=2t$. La courbe en tiret\'es noirs correspond \`a la courbe moyenne de la figure \ref{fig:Energie-1} avec ses barres d'erreur.}
\label{fig:decenergU2}
\end{figure}
\begin{figure}
\resizebox{0.85\columnwidth}{!}{%
  \includegraphics{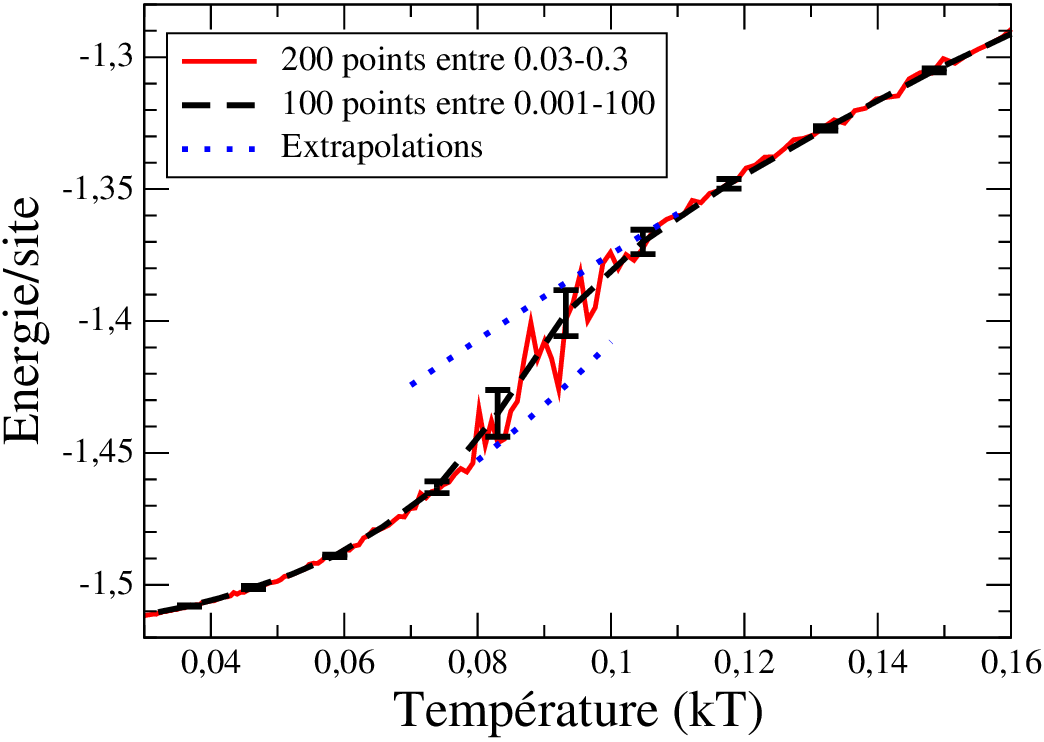}
}
\caption{(Couleur en ligne) Courbes d'\'energie pour $U=3t$. La courbe en tiret\'es noirs correspond \`a la courbe moyenne de la figure \ref{fig:Energie-1} avec ses barres d'erreur.}
\label{fig:decenergU3}
\end{figure}
\begin{figure}
\resizebox{0.9\columnwidth}{!}{%
  \includegraphics{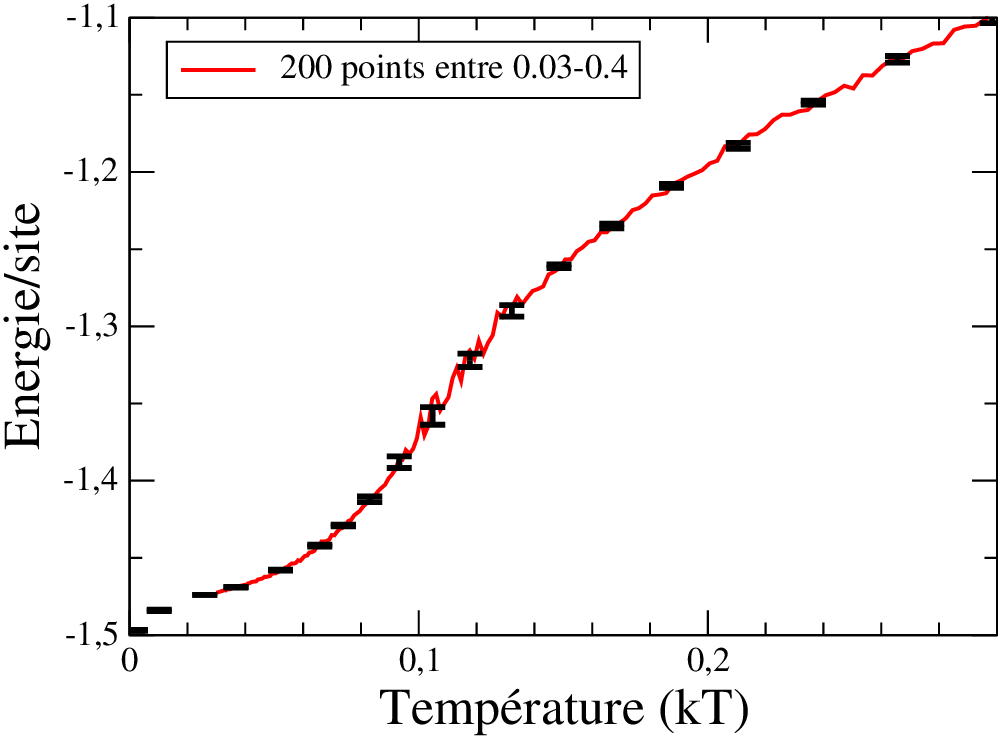}
}
\caption{(Couleur en ligne) Courbes d'\'energie pour $U=3.3t$. Les barres d'erreur correspondent \`a la courbe moyenne de la figure \ref{fig:Energie-1}.}
\label{fig:decenergU3.3}
\end{figure}
\begin{figure}
\resizebox{0.9\columnwidth}{!}{%
  \includegraphics{figure-17.eps}
}
\caption{(Couleur en ligne) Courbes d'\'energie pour $U=3.5t$. Les barres d'erreur correspondent \`a la courbe moyenne de la figure \ref{fig:Energie-1}.}
\label{fig:decenergU3.5}
\end{figure}
\begin{figure}
\resizebox{0.9\columnwidth}{!}{%
  \includegraphics{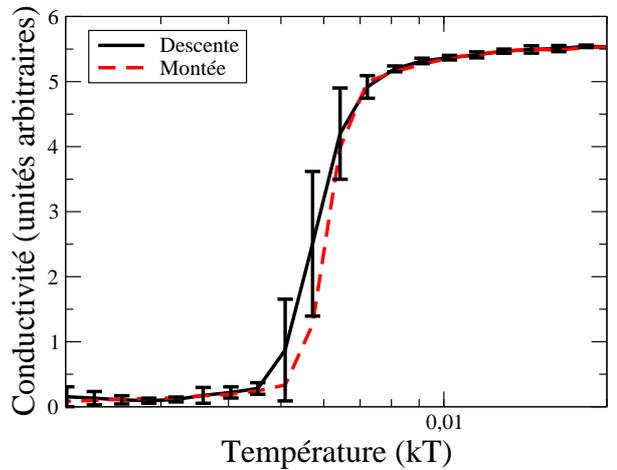}
}
\caption{(Couleur en ligne) Courbes de conductivit\'e en descente et en mont\'ee de temp\'erature pour $U=0.5t$. Les barres d'erreur correspondent \`a la courbe en descente de temp\'erature.}
\label{fig:hystconduct0.5}
\end{figure}
\subsection{Ph\'enom\`ene d'hyst\'er\'esis}
Les figures \ref{fig:hystconduct0.5} \`a \ref{fig:hystconduct3.5} pr\'esentent les courbes de conductivit\'e moyennes calcul\'ees en descente et en mont\'ee de temp\'erature. Ces courbes moyennes font appara\^itre un ph\'enom\`ene d'hyst\'er\'esis entre la mont\'ee et la descente de temp\'erature dans la zone de transition, pour les valeurs de $U$ inf\'erieures \`a $3.5t$. Bien qu'il s'agisse de courbes moyennes et que les courbes en mont\'ee de temp\'erature soient comprises dans les barres d'erreur des courbes trac\'ees en descente, la r\'ep\'etition syst\'ematique de ce ph\'enom\`ene pour toutes les valeurs de $U<3.5t$ indique que ce n'est pas un ph\'enom\`ene al\'eatoire. Cette mise en \'evidence d'un ph\'enom\`ene d'hyst\'er\'esis doit, cependant, \^etre consid\'er\'ee avec prudence. En effet, conform\'ement au principe des m\'ethodes de Monte-Carlo, les \'etats g\'en\'er\'es \`a une temp\'earture donn\'ee ne doivent pas \^etre fonction des \'etats pr\'ec\'edents.
\begin{figure}
\resizebox{0.9\columnwidth}{!}{%
  \includegraphics{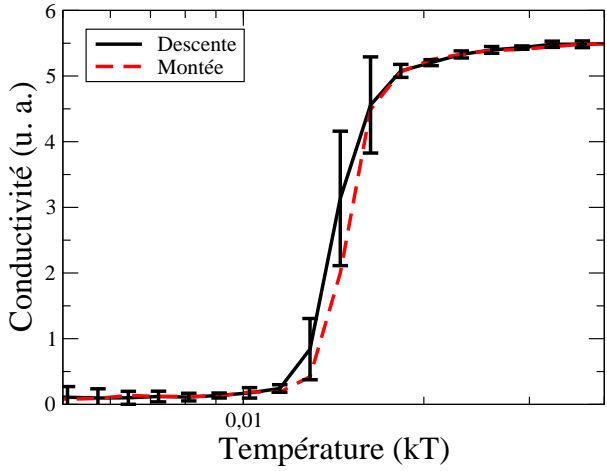}
}
\caption{(Couleur en ligne) Courbes de conductivit\'e en descente et en mont\'ee de temp\'erature pour $U=1t$. Les barres d'erreur correspondent \`a la courbe en descente de temp\'erature.}
\label{fig:hystconduct1}
\end{figure}
\begin{figure}
\resizebox{0.9\columnwidth}{!}{%
  \includegraphics{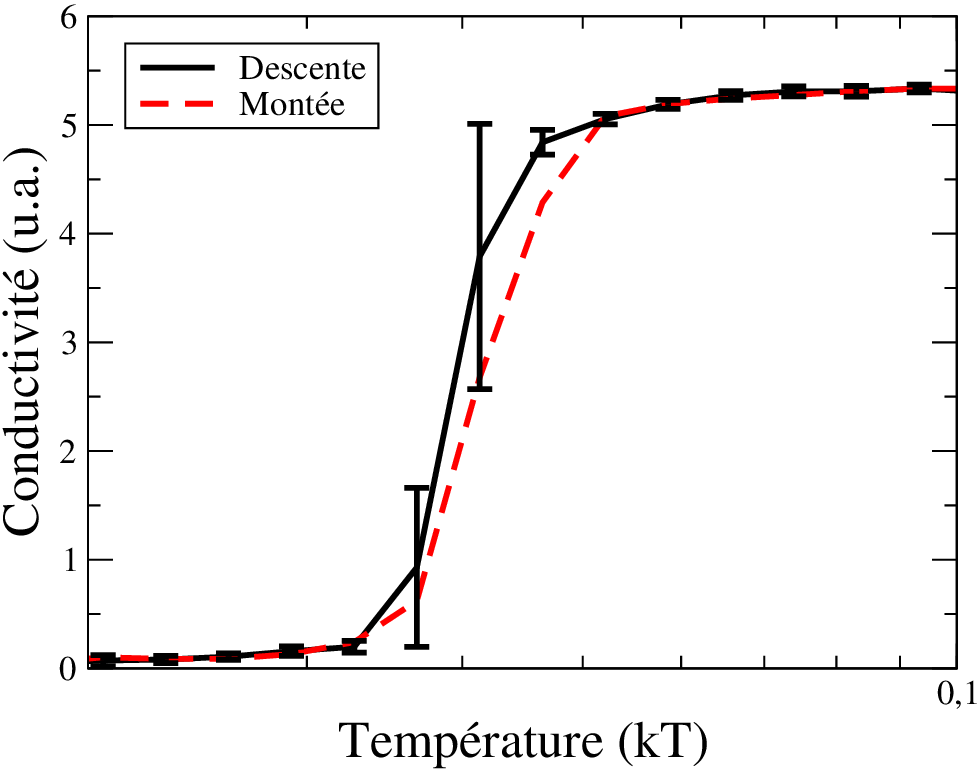}
}
\caption{(Couleur en ligne) Courbes de conductivit\'e en descente et en mont\'ee de temp\'erature pour $U=2t$. Les barres d'erreur correspondent \`a la courbe en descente de temp\'erature.}
\label{fig:hystconduct2}
\end{figure}
\begin{figure}
\resizebox{0.9\columnwidth}{!}{%
  \includegraphics{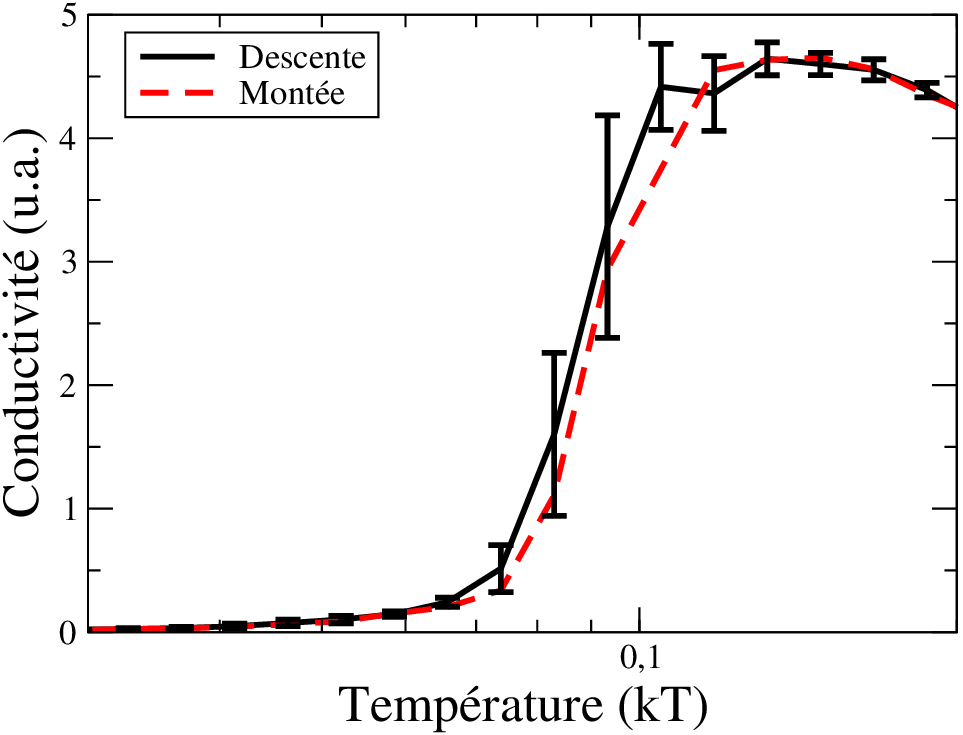}
}
\caption{(Couleur en ligne) Courbes de conductivit\'e en descente et en mont\'ee de temp\'erature pour $U=3t$. Les barres d'erreur correspondent \`a la courbe en descente de temp\'erature.}
\label{fig:hystconduct3}
\end{figure}
\begin{figure}
\resizebox{0.9\columnwidth}{!}{%
  \includegraphics{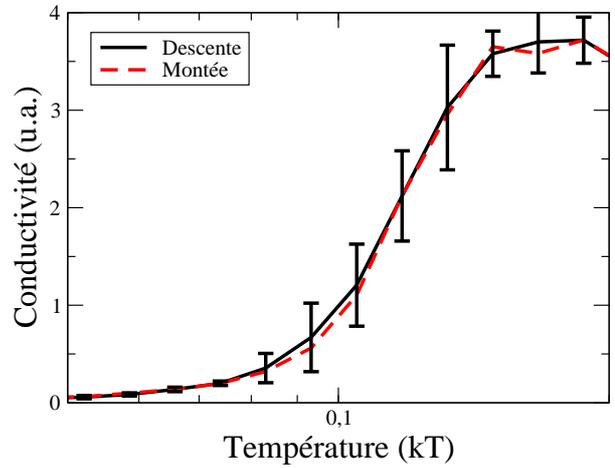}
}
\caption{(Couleur en ligne) Courbes de conductivit\'e en descente et en mont\'ee de temp\'erature pour $U=3.3t$. Les barres d'erreur correspondent \`a la courbe en descente de temp\'erature.}
\label{fig:hystconduct3.3}
\end{figure}
\begin{figure}
\resizebox{0.9\columnwidth}{!}{%
  \includegraphics{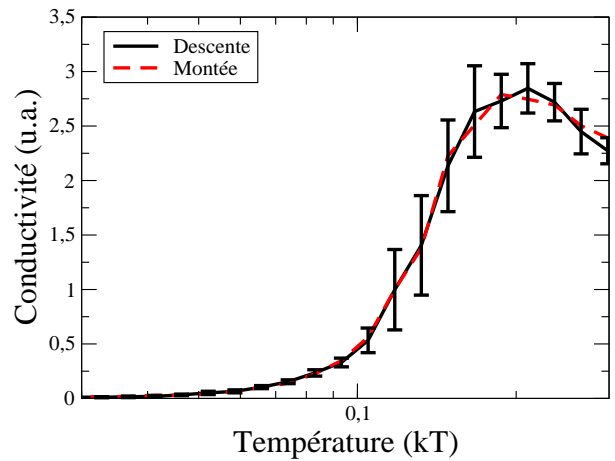}
}
\caption{(Couleur en ligne) Courbes de conductivit\'e en descente et en mont\'ee de temp\'erature pour $U=3.5t$. Les barres d'erreur correspondent \`a la courbe en descente de temp\'erature.}
\label{fig:hystconduct3.5}
\end{figure}
Ce ph\'enom\`ene d'hyst\'er\'esis appara\^it \'egalement sur les courbes du facteur d'antiferromagn\'etisme.
\subsection{R\'esultats en fonction de U/t}
Comme il a \'et\'e indiqu\'e pr\'ec\'edemment, les simulations ont \'et\'e r\'ealis\'ees en programmant des cycles de temp\'erature. Les r\'esultats obtenus ont permis de tracer les variations de diff\'erentes grandeurs en fonction de la temp\'erature. Ceci est parfaitement adapt\'e pour mettre en \'evidence des transitions de phases induites par les variations de temp\'erature. Par contre, la transition de Mott attendue doit \^etre, principalement, produite par les variations de l'interaction coulombienne $U/t$. Il est donc souhaitable de tracer l'\'evolution des diff\'erentes grandeurs en fonction de $U/t$. Ceci est indispensable si la ligne de transition est parall\`ele \`a l'axe des temp\'eratures dans le plan $\left( U/t,kT/t\right) $, c'est \`a dire si elle n'est pas travers\'ee en programmant des cycles de temp\'eratures. Les r\'esultats des simulations ont donc \'et\'e trait\'es pour obtenir les courbes souhait\'ees. Les isothermes d'\'energie , de chaleur sp\'ecifique, de conductivit\'e, du facteur d'antiferromagn\'etisme et de double occupation sont pr\'esent\'ees dans les figures \ref{fig:energieU} \`a \ref{fig:doubloccupU}.
\begin{figure}
\resizebox{0.9\columnwidth}{!}{%
  \includegraphics{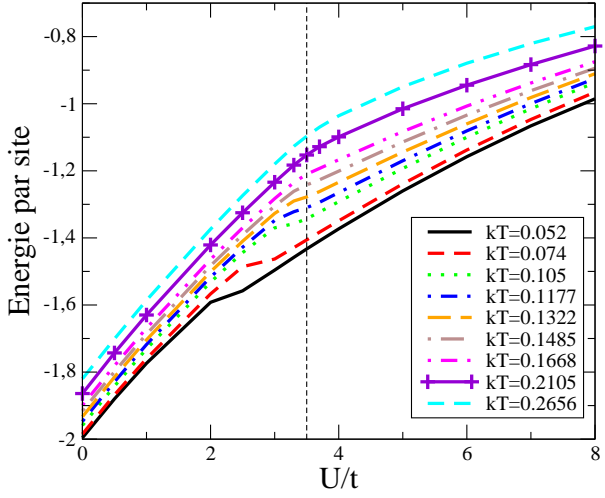}
}
\caption{(Couleur en ligne) Isothermes d'\'energie en fonction de $U/t$. Seuls les points sur l'isotherme $kT=0.2105$ ont \'et\'e report\'es. Les points sur les autres courbes correspondent aux m\^emes valeurs de $U/t$.}
\label{fig:energieU}
\end{figure}
\begin{figure}
\resizebox{0.9\columnwidth}{!}{%
  \includegraphics{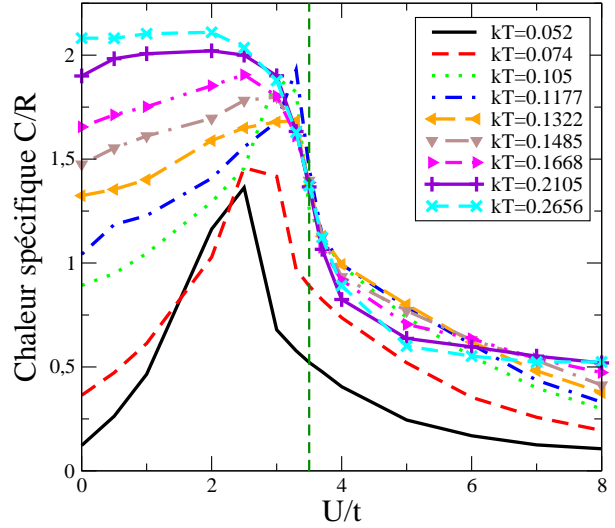}
}
\caption{(Couleur en ligne) Isothermes de chaleur sp\'ecifique en fonction de $U/t$. Pour $0.2656 \geq kT \geq 0.13t$ les courbes se superposent dans la zone de transition.}
\label{fig:chaleurspecU}
\end{figure}
\begin{figure}
\resizebox{0.9\columnwidth}{!}{%
  \includegraphics{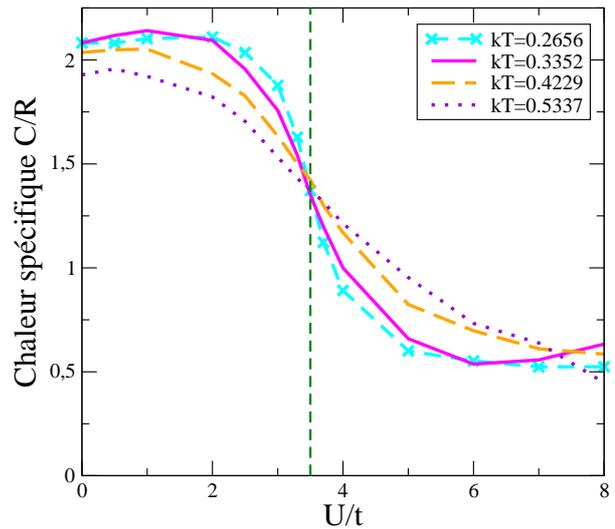}
}
\caption{(Couleur en ligne) Isothermes de chaleur sp\'ecifique en fonction de $U/t$ pour les hautes temp\'eratures $\left( kT \geq 0.2656t \right) $.}
\label{fig:chaleurspecU-2}
\end{figure}
\begin{figure}
\resizebox{0.9\columnwidth}{!}{%
  \includegraphics{figure-27.eps}
}
\caption{(Couleur en ligne) Isothermes de conductivit\'e en fonction de $U/t$. Seuls les points sur l'isotherme $kT=0.2105$ ont \'et\'e report\'es. Les points sur les autres courbes correspondent aux m\^emes valeurs de $U/t$.}
\label{fig:conductivU}
\end{figure}
\begin{figure}
\resizebox{0.9\columnwidth}{!}{%
  \includegraphics{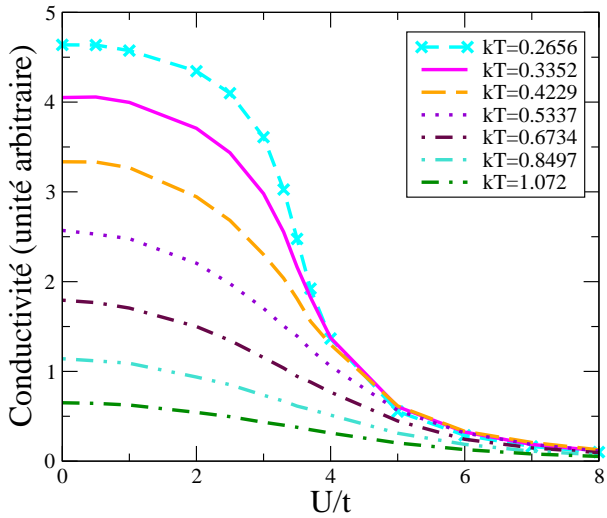}
}
\caption{(Couleur en ligne) Isothermes de conductivit\'e en fonction de $U/t$ pour les hautes temp\'eratures.}
\label{fig:conductivU-3}
\end{figure}
\begin{figure}
\resizebox{0.9\columnwidth}{!}{%
  \includegraphics{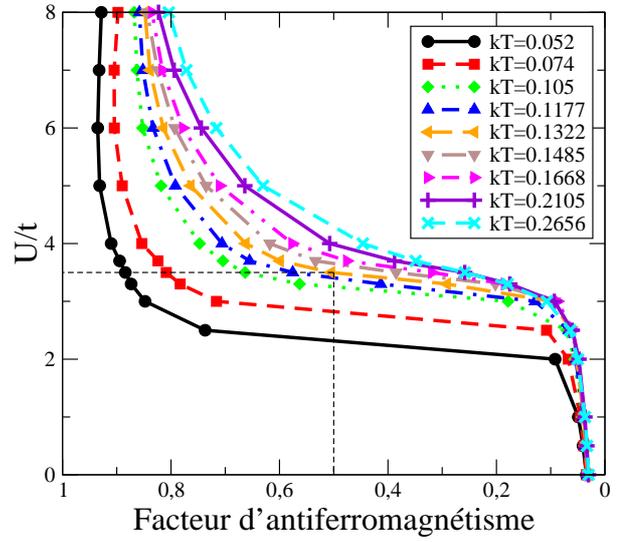}
}
\caption{(Couleur en ligne) Isothermes du facteur d'antiferromagn\'etisme en fonction de $U/t$.}
\label{fig:antiferroU}
\end{figure}
\begin{figure}
\resizebox{0.9\columnwidth}{!}{%
  \includegraphics{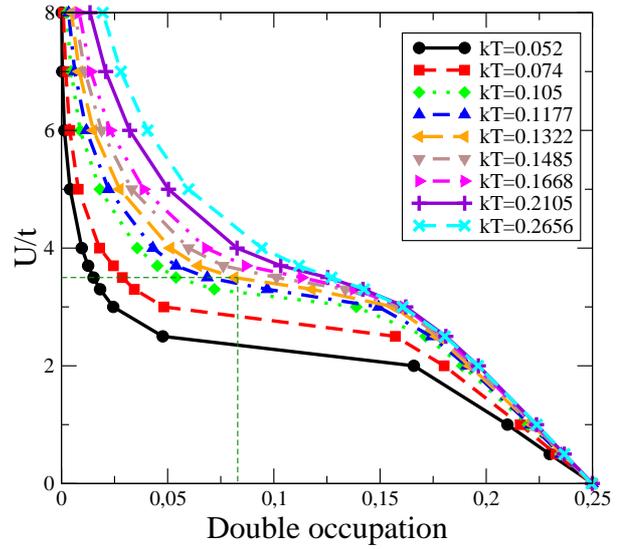}
}
\caption{(Couleur en ligne) Isothermes de double occupation $D$ en fonction de $U/t$.}
\label{fig:doubloccupU}
\end{figure}
\section{Discussion}
\label{sec:discussion}
\subsection{Transition m\'etal-isolant du premier ordre}
Les courbes de chaleur sp\'ecifique de la figure \ref{fig:Chalspec-1}, de conductivit\'e des figures \ref{fig:Conductiv} et \ref{fig:conduct-lin} et du facteur d'antiferromagn\'etisme des figures \ref{fig:factantiferro} et \ref{fig:factantifer-lin} font clairement appara\^itre une transition de phase, quand la temp\'erature varie, pour les valeurs de l'interaction coulombienne telles que $U \leq 3.5t$.  La phase basse temp\'erature est une phase isolante antiferromagn\'etique. La phase haute temp\'erature est une phase conductrice sans ordre magn\'etique. Les courbes de conductivit\'e \`a basse temp\'erature de la figure \ref{fig:conduct-lin} montrent que cette phase a un comportement m\'etallique. La transition est donc une transition m\'etal-isolant concomittante \`a une transition paramagn\'etique-antiferromagn\'etique.
Les courbes moyennes des figures \ref{fig:Energie-1} \`a \ref{fig:paires} ne permettent pas de d\'eterminer avec certitude la nature de cette transition. Par contre, les figures \ref{fig:decenergU0.5} \`a \ref{fig:decenergU3.5} montrent clairement que les courbes d'\'energie moyenne comportent un d\'ecrochement caract\'eristique des transitions du premier ordre. Ce d\'ecrochement est confirm\'e par les fluctuations des courbes avec $200$ points de mesure des figures \ref{fig:decenergU0.5} \`a \ref{fig:decenergU3.3}. En effet, ces courbes s'interpr\`etent en consid\'erant que, dans les plages des temp\'eratures de transition, les \'etats du syst\`eme correspondent \`a deux puits de potentiels. Ainsi, les \'etats successifs g\'en\'er\'es par l'algorithme peuvent rester ``bloqu\'es'' longtemps dans un des puits de potentiel. Ceci justifie que les valeurs de l'\'energie fluctuent entre deux valeurs limites. Il y a donc un ph\'enom\`ene de m\'etastabilit\'e entre deux \'etats macroscopiques possibles du syst\`eme. Ce ph\'enom\`ene de m\'etastabilit\'e et le ph\'enom\`ene d'hyst\'er\'esis observ\'e sur les courbes de conductivit\'e sont \'egalement des propri\'et\'es caract\'eristiques des transitions du premier ordre. On peut donc conclure, sans aucun doute, que la transition est du premier ordre. On en d\'eduit que les pics de chaleur sp\'ecifique ne sont pas correctes. De m\^eme, les valeurs moyennes des autres grandeurs ne sont pas exactes dans les plages de temp\'eratures o\`u il y a le ph\'enom\`ene de m\'etastabilit\'e. En effet, dans la zone de transition, les configurations g\'en\'er\'ees par l'algorithme qui sont retenues pour calculer les diff\'erentes grandeurs, sont des \'etats microscopiques des deux phases. C'est \`a dire que le syst\`eme ``oscille'' entre les deux phases. Ce ph\'enom\`ene de m\'etastabilit\'e signifie que le syst\`eme n'est pas \`a l'\'equilibre thermodynamique, donc la relation \ref{specifheat1} qui permet de calculer la chaleur sp\'ecifique ne peut pas \^etre utilis\'ee. La singularit\'e math\'ematique attendue sur la courbe de chaleur sp\'ecifique d'une transition du premier ordre est un saut ou d\'ecrochement comme pour l'\'energie.
\subsection{Point critique}
Comme cela a \'et\'e remarqu\'e pr\'ec\'edemment, un changement de comportement se produit pour $U\simeq3.5t$. Cette valeur est la valeur limite d'une ligne de transition du premier ordre dans le plan $\left( U/t, kT/t\right) $. La temp\'erature correspondante, \`a mi-hauteur du pic, est environ $kT/t\simeq 0.13$. Ces deux valeurs sont les coordonn\'ees du point critique qui termine la ligne de transition du premier ordre. Les temp\'eratures des points de cette ligne de transition sont \'egalement d\'etermin\'ees \`a mi-hauteur des pics. Le r\'eseau d'isothermes de la figure \ref{fig:antiferroU} confirme qu'il s'agit bien d'un point critique analogue \`a celui observ\'e pour les transitons liquide-gaz.\\
On remarque, sur les figures \ref{fig:antiferroU} et \ref{fig:doubloccupU}, que les paliers des isothermes telles que $kT/t< 0.1322$ ne sont pas parall\`eles \`a l'axe des ``x''. Ceci est d\^u \`a la m\'ethode de Monte-Carlo. En effet, comme cela a \'et\'e indiqu\'e pr\'ec\'edemment, dans une zone de transition du premier ordre cette m\'ethode g\'en\`ere des  \'etats des deux phases possibles, ainsi les moyennes obtenues ne correspondent pas \`a une seule phase. On obtient alors des valeurs interm\'ediaires entre les valeurs relatives \`a chaque phase. Il faudrait, pour contourner ce probl\`eme, s\'electionner les configurations retenues en ne gardant que celles correspondant \`a une seule phase, c'est \`a dire \`a un puits de potentiel. Malheureusement, il appara\^it difficile de d\'eterminer \`a quelle phase correspond une configuration g\'en\'er\'ee. Aussi, quels que soient les nombres de points report\'es sur les courbes, les paliers ne seront jamais parfaitement parall\`eles \`a l'axe des ``x''.
\subsection{Transition de Mott}
Alors que les r\'eseaux d'isothermes du facteur d'antiferromagn\'etisme de la figure \ref{fig:antiferroU} et du facteur de double occupation de la figure \ref{fig:doubloccupU} confirment l'existence du point critique ($U/t\simeq3.5$, $kT/t\simeq0.13$), les isothermes de conductivit\'e et de chaleur sp\'ecifique des figures \ref{fig:conductivU} et \ref{fig:chaleurspecU} font appa\^itre un saut de conductivit\'e associ\'e \`a un saut de chaleur sp\'ecifique pour les temp\'eratures $0.2656t \geq kT \geq 0.13t$. Ce saut se produit pour la valeur constante $U/t \simeq 3.5$, \`a mi-hauteur du d\'ecrochement. Les figures \ref{fig:chaleurspecU-2} et \ref{fig:conductivU-3} montrent que pour les temp\'eratures $kT > 0.2656$ ce saut dispara\^it, les courbes pr\'esentent alors un point d'inflexion. Cette discontinuit\'e doit correspondre \`a une transition m\'etal-isolant de Mott. La pr\'esence de cette transition est \'egalement observ\'ee sur les isothermes d'\'energie de la figure \ref{fig:energieU}. En effet, ces courbes pr\'esentent un point anguleux pour la valeur de l'interaction coulombienne $U/t\simeq 3.5$. Cependant, les r\'esultats ne permettent pas de d\'eterminer de fa\c{c}on certaine la nature de cette transition. Etant donn\'e le changement de comportement observ\'e sur les courbes des figures \ref{fig:chaleurspecU-2} et \ref{fig:conductivU-3} pour les temp\'eratures $kT\geq 0.2656t$, on peut supposer que la transition est encore du premier ordre dans l'intervalle de temp\'erature $0.1322-0.2656$. Le point ($U/t=3.5$, $kT/t=0.2656$) serait alors un point critique sur la ligne de transition de Mott. Dans cette hypoth\`ese, les deux lignes de transition du premier ordre doivent se rejoindre en un point triple proche du point critique de la transition isolant antiferromagn\'etique-conducteur paramagn\'etique.
\subsection{Diagramme de phase}
On peut consid\'erer que la valeur moiti\'e de la conductivit\'e maximum est la valeur limite entre le comportement conducteur et le comportement mauvais conducteur. Ainsi pour des valeurs de la conductivit\'e l\'eg\`erement sup\'erieures le syst\`eme se comporte comme un conducteur tandis que pour des valeurs l\'eg\`erement inf\'erieures le syst\`eme peut \^etre qualifi\'e de mauvais conducteur. La ligne en pointill\'es (rouge) de la figure \ref{fig:diagram-3} correspond \`a cette limite de comportement.\\
Les courbes de conductivit\'e montrent que le syst\`eme est isolant et paramagn\'etique, quel que soit $U$, quand $kT>2t$. Cependant, les courbes de la figure \ref{fig:paires} montrent que ces \'etats paramagn\'etiques sont caract\'eris\'es par des nombres de double occupation diff\'erents. De m\^eme on constate que le syst\`eme est isolant, quelle que soit la temp\'erature pour $U \gtrsim 8t$, mais le syst\`eme est antiferromagn\'etique \`a basse temp\'erature et devient progressivement paramagn\'etique quand la temp\'erature augmente. Ce passage isolant antiferromagn\'etique-isolant paramagn\'etique se produit sans qu'il y ait de transition, au sens thermodynamique, d\'etectable sur toutes les grandeurs calcul\'ees dans nos simulations. Ceci est peut \^etre d\^u \`a la petite taille du r\'eseau.\\
Comme pour le changement de comportement conducteur-mauvais conducteur, on peut consid\'erer que le syst\`eme pr\'esente une tendance au ferromagn\'etisme pour $f_{A}>0.5$ et au paramagn\'etisme pour $f_{A}<0.5$. Ceci permet de tracer une ligne de ``cross-over'' entre les deux phases magn\'etiques.\\
Selon les hypoth\`eses pr\'ec\'edentes le diagramme de phase dans le plan $\left( U/t,kT/t\right) $ comprend, donc, une premi\`ere ligne de transition du premier ordre pour $kT/t<0.13$ et $U/t<3.5$ et une deuxi\`eme ligne de transition du premier ordre pour $0.2656 \geq kT \geq 0.13$, cette ligne \'etant parall\`ele \`a l'axe des temp\'eratures. Ces deux lignes se rejoignent en un point triple. Les deux lignes de ``cross-over'' d\'efinies pr\'ec\'edemment peuvent compl\'eter le diagramme. Toutes ces consid\'erations permettent de tracer le diagramme de phase de la figure \ref{fig:diagram-3}.
%
%
\begin{figure}
\resizebox{0.9\columnwidth}{!}{%
  \includegraphics{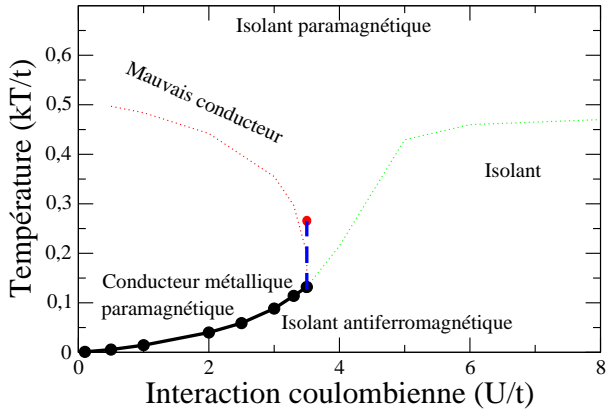}
}
\caption{(Couleur en ligne) Allure g\'en\'erale du diagramme de phase du mod\`ele de Hubbard 2D \`a demi-remplissage. Les lignes en pointill\'es (rouge) et (vert) sont des lignes de ``cross-over''. La ligne en trait plein (noir) est la ligne de transition du premier ordre isolant antiferromagn\'etique-conducteur paramagn\'etique. La ligne en tiret\'es (bleu) correspond \`a la transition de Mott.}
\label{fig:diagram-3}
\end{figure}
\subsection{Comparaison avec quelques r\'esultats de la litt\'erature}
La transition basse temp\'erature m\'etal paramagn\'etique - isolant antiferromagn\'etique a effectivement \'et\'e observ\'ee exp\'erimentalement sur un certain nombre de compos\'es, notamment $V_{2}0_{3}$ \cite{Foex,Limelette,kuwamoto}. Cette transition est effectivement du premier ordre. Les r\'esultats obtenus en faisant varier la pression mettent en \'evidence une transition m\'etal-isolant \`a haute temp\'erature qui correspond \`a la transition de Mott. Cette transition est \'egalement une transition du premier ordre. Les r\'esultats de nos simulations sont donc en bon accord qualitatif avec l'exp\'erience.\\
Les diagrammes de phases exp\'erimentaux font \'egalement appara\^itre une ligne de transition entre la phase isolante antiferromagn\'etique et la phase isolante paramagn\'etique. Cette transition n'a pas \'et\'e d\'etect\'ee dans nos simulations. Ces diff\'erences peuvent \^etre, \'eventuellement, justifi\'ees en consid\'erant que le mod\`ele \'etudi\'e est un mod\`ele 2D alors que les r\'esultats exp\'erimentaux concernent des mat\'eriaux 3D. De m\^eme, l'influence de l'interaction coulombienne $U$ n'est certainement pas totalement \'equivalente \`a celle de la pression.\\
De nombreux r\'esultats de simulations et d'\'etudes th\'eoriques du mod\`ele de Hubbard 2D ont \'et\'e publi\'es. Nous ne consid\`ererons que quelques un de ces articles qui pr\'esentent les courbes des m\^emes grandeurs que celles calcul\'ees dans nos simulations. De fa\c{c}on g\'en\'erale, on remarque que l'allure de notre diagramme de phases est proche de celle des diagrammes obtenus par d'autres m\'ethodes, pour des mod\`eles non frustr\'es. En particulier,
on constate que le point critique que nous avons d\'etermin\'e correspond au point triple du diagramme de phases de la r\'ef\'erence \cite{pruschke}. Les courbes d'\'energie et de chaleurs sp\'ecifiques ont \'egalement une allure g\'en\'erale tr\`es proches de celles pr\'esent\'ees dans les r\'ef\'erences \cite{vries,Georges,Duffy,Paiva,Aryanpour}. Mais les r\'esultats de ces r\'ef\'erences ne font pas appara\^itre les sauts de l'\'energie qui permettent de d\'eterminer la nature de la transition. On constate \'egalement que les courbes du carr\'e de l'aimantation locale $\left\langle  m_{z} ^{2} \right\rangle $ de la figure \ref{fig:mz2-lin} ont la m\^eme allure que celles pr\'esent\'ees dans la r\'ef\'erence \cite{Aryanpour}. Cependant, une diff\'erence appara\^it. En effet nos courbes pr\'esentent, \`a basse temp\'erature, des variations qui correspondent \`a la transition m\'etal-isolant du premier ordre alors que cette transition n'est pas d\'etect\'ee dans la r\'ef\'erence.\\
Gröber et al. \cite{Grober} ont r\'ealis\'e des simulations de Mont\'e-Carlo quantique du mod\`ele 2D de Hubbard $(8\times8)$ en faisant varier la valeur de l'interaction coulombienne $U$ pour la temp\'erature $T=0.33t$. Ils ont mis en \'evidence une transition m\'etal paramagn\'etique-isolant pour la valeur critique $U_c\approx4t$. Cette valeur critique est  proche de la valeur $U\simeq3.5t$ que nous avons obtenue.
\section{Conclusion}
\label{sec:conclusion}
Comme toute m\'ethode num\'erique, la m\'ethode employ\'ee pour nos simulations utilise des approximations  qui permettent de r\'eduire consid\'erablement les calculs. La validit\'e des r\'esultats obtenus peut \^etre mise en doute \`a cause de ces approximations. Mais la coh\'erence des r\'esultats et le bon accord qualitatif avec un certain nombre de r\'esultats obtenus avec d'autres m\'ethodes num\'eriques, avec les pr\'edictions th\'eoriques et les r\'esultats exp\'erimentaux permet de consid\'erer que ces r\'esultats sont significatifs.\\
Le Mod\`ele de Hubbard 2D \'etudi\'e fait appara\^itre deux transitions: une transition du premier ordre m\'etal paramagn\'etique - isolant antiferromagn\'etique \`a basse temp\'erature et la transition de Mott m\'etal paramagn\'etique-isolant paramagn\'etique pour $0.2656 \geq kT/t \geq 0.13$. Par contre, le passage isolant antiferromagn\'etique-isolant paramagn\'etique se produit sans transition de phase d\'etectable sur les grandeurs \'etudi\'ees. Il est possible que l'effet de taille ne permette pas d'observer tous les ``d\'etails'' du comportement du mod\`ele qui semble \^etre compliqu\'e autour du point critique.
%
%

\end{twocolumn}
\end{document}